\documentclass[pre,twocolumn,showpacs,amsmath,superscriptaddress,amssymb,floatfix]{revtex4}
\usepackage[dvips]{graphicx}
\usepackage{dcolumn}

\begin{document}
\title{Diffusive Dynamics of Water inside Hydrophobic Carbon Micropores Studied by Neutron Spectroscopy and Molecular  Dynamics Simulation}
\author{S.O. Diallo}
\email{omardiallos@ornl.gov}
\affiliation{Quantum Condensed Matter Division,  Oak Ridge National Laboratory, Oak Ridge, Tennessee 37831, USA}
\author{L. Vlcek}
\affiliation{Chemical Sciences Division, Oak Ridge, National Laboratory, Tennessee 37831, USA }
\author{E. Mamontov}
\affiliation{Chemical and Engineering Science Division,  Oak Ridge National Laboratory, Oak Ridge, Tennessee 37831, USA}
\author{J.K. Keum}
\affiliation{Instrument and Source Division, Oak Ridge National Laboratory, Oak Ridge, Tennessee 37831, USA}
\author{Jihua Chen}
\affiliation{Center for Nanophase Materials Sciences, Oak Ridge National Laboratory, Oak Ridge, Tennessee 37831, USA}
\author{J.S. Hayes Jr.}
\affiliation{American Technical Trading, Incorporated, P.O. Box 273, Pleasantville, New York  10570, USA}
\author{A.A. Chialvo}
\email{chialvoaa@ornl.gov}
\affiliation{Chemical Sciences Division, Oak Ridge, National Laboratory, Tennessee 37831, USA }


\begin{abstract}
When water molecules are confined to nanoscale spacings, such as in the nanometer size pores of activated carbon fiber (ACF), their freezing point gets suppressed down to very low temperatures ($\sim$ 150 K), leading to a metastable liquid state with remarkable physical properties.  We have investigated  the ambient pressure diffusive dynamics of water in microporous  Kynol\texttrademark ACF-10 (average pore size $\sim$11.6 {\AA}, with primarily slit-like pores) from temperature $T=$ 280 K in its stable liquid state down to $T=$ 230 K into the metastable supercooled phase.  The observed characteristic relaxation times and diffusion coefficients are found to be respectively higher and lower than those in bulk water, indicating a slowing down of the water mobility with decreasing temperature. The observed temperature-dependent average relaxation time $\langle\tau\rangle$ when compared to previous findings indicate that it is the size of the confining pores - not their shape - that primarily affects the dynamics of water for pore sizes larger than 10 {\AA}. The experimental observations are compared to complementary molecular dynamics  simulations of a model system, in which we studied the diffusion of water within the 11.6 {\AA}  gap of two parallel graphene sheets. We find generally a reasonable agreement between the observed and calculated relaxation times at the low momentum transfer $Q$ ($Q\le 0.9$ \AA${^{-1}}$).  At high $Q$ however, where localized dynamics becomes relevant, this ideal system does not satisfactorily reproduce the measurements. Consequently,  the simulations are compared to the experiments at low $Q$, where the two can be best reconciled. The best agreement is obtained for the diffusion parameter $D$ associated with the hydrogen-site when a representative stretched exponential function, rather than the standard bi-modal exponential model, is used to parameterize the self-correlation function  $I(Q,t)$ . 
\end{abstract}

\maketitle

\date{\today}

\section{Introduction}\label{sec0}
Fluids confined to nano-cavities, or restricted to surfaces are pervasive in nature \cite{Stanley:99}.  Such nano-confinement often gives rise to unusual physical properties, that are distinct from those in the bulk liquid \cite{Chaplin:09,Sliwinska:12}. In the case of water, nano-confinement suppresses crystallization altogether,  modifying the water phase diagram and leading to a metastable supercooled liquid.   Nano-porous carbon compounds are perfect examples of confining media that have inspired research on supercooled water \cite{Zhang:09a,Kolesnikov:04}, and that continue to stimulate applied research for developing for example high-performing membranes for water desalination, nanofluidic devices, and alternative drug delivery methods\cite{Striolo:07}.  

The dynamical properties of supercooled water constrained in cylindrical pores have been extensively investigated with neutron scattering using various hydrophilic silica materials  such as MCM-14 or FSM-12 \cite{Faraone:04,Chen:06a,Liu:06,Takahara:05,Chu:10,Diallo:12}, and some hydrophobic carbon systems such as carbide-derived carbon (CDC)\cite{Chathoh:11}, single-wall (SWNT)\cite{Mamontov:06}, and double-wall nanotubes (DWNT) \cite{Mamontov:06,Chu:07}.   In real physical systems however, water confinement occurs under various spatial and geometrical restrictions. Understanding not only  the effects of varying pore size on the structural and dynamical properties of water, but also the influence of different pore shapes is thus of great scientific importance.

The present study aims to understand how supercooled water behaves when it is confined inside a hydrophobic pore structure that is non-cylindrical, and thereby evaluate the effect of geometry on the water dynamics. Specifically, we here examine  the characteristic diffusive dynamics of water confined in the narrow pores of Activated Carbon Fiber (ACF-10) using Quasi-Elastic Neutron Scattering (QENS).   The observed average characteristic relaxation $\langle \tau\rangle$  and diffusion coefficient $D$  (associated with the hydrogen sites of water) indicate a diffusion that is slower than in the bulk liquid, with an Arrhenius behavior with characteristic energy  barriers $E_A$ in the range $14\le E_A\le29$ kJ/mol, depending on the fit model of various characteristic parameters.  Our molecular dynamics  simulation of water  confined between two parallel flat graphene-surfaces, while not fully representative of the real physical system, shows qualitatively the same systematic trend with temperature, with an $E_A\simeq19$ kJ/mol. By comparing the present findings with previous experimental observations, we conclude that at any given thermodynamic condition, it is primarily the confining pore dimension, and not its geometry, that dictates the dynamical behavior of confined water. The measured and calculated intrinsic intermediate scattering function  $I_{in}(Q,t)$ (or self-correlation function) at ambient conditions could be satisfactorily reconciled at the low momentum transfer $Q$ ($\le$0.9 \AA${^{-1}}$), where localized dynamics are not as relevant. 

\begin{figure}
\includegraphics[width=1.00\linewidth]{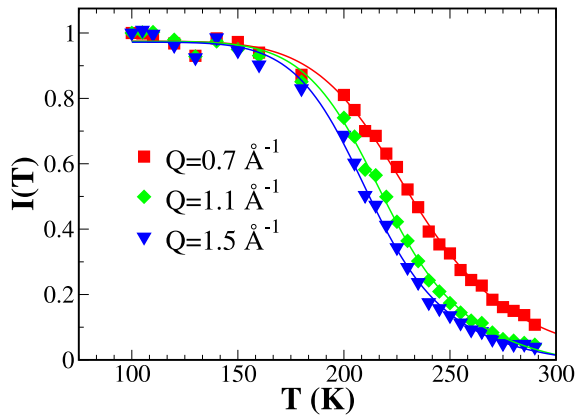}
\caption{(Color online) Temperature dependence of the integrated elastic scattering $I(T)$ of water confined in ACF-10, at a few selected momentum transfers $Q$. Data are normalized with respect to the lowest temperature $(T_0=100)$ K, as indicated. The onset of observable diffusive dynamics on BASIS is marked by the departure from a monotonically varying slope in $I(T)$ at low temperatures to a more rapidly changing slope at the higher temperatures. The  solid lines are model fits to data, corresponding to an activation energy $E_A=$14-15 kJ/mol.}
\label{fig.1}
\end{figure}

\begin{figure}
\includegraphics[width=1.\linewidth]{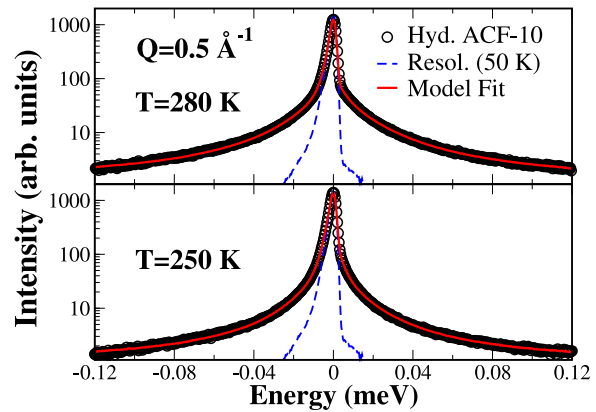}
\caption{(Color online) Representative spectrum of water in AFC-10  at $Q=0.5$ \AA$^{-1}$, as observed on BASIS, plotted with the measured BASIS resolution function at 50 K (blue dashed line). Data at 280 K (top panel) and at 250K (bottom panel) are shown. The open circles represent the total observed signal from the water wetted ACF-10 sample in the Al holder. The red solid lines are model fits to the experimental data, as discussed in the text.}
\label{fig.2}
\end{figure}

\section{Neutron Experiments} \label{sec:exp}
In this section, we describe the ACF-10 sample, and the various steps involved in preparing for the neutron measurements. We also discuss the actual measurements performed on the backscattering spectrometer (BASIS) at Oak Ridge National Laboratory (ORNL), USA \cite{Mamontov:11}. This inverted geometry spectrometer has an excellent energy resolution of 3.5 $\mu$eV (2 $\times$ Half-Width-at-Half-Maximum = 2  $\times \Gamma_r$) at the elastic line, and covers a momentum $Q$ and energy $E$ transfer range, respectively 0.3 $\le Q\le$ 1.9 {\AA}$^{-1}$, and  -0.12 $\le E\le$ 0.12  meV when operated  with incident neutron beam with wavelength $\lambda_i=$ 6.4 {\AA} ($\Delta\lambda_i=\pm0.5$ \AA), corresponding to an energy bandwidth of 1.71 $\le E_i \le$ 2.35 meV.

\subsection{Sample Details}
The Kynol\texttrademark $^*$ ACF-10 (ACF-1603-10) sample was provided to us by American Technical Trading (ATT), NY.  With its human hair-like appearance, this activated carbon fiber  is reported  to have an \lq average' length of 2-4 mm, and diameter $\sim$ 10 $\mu$m, which is consistent with our transmission electron microscopy conducted at the Oak Ridge National Laboratory's Center of Nanophase Materials Sciences.  Previously reported scanning tunneling microscopy (STM) of ACF-10 \cite{Kaneko:92a,Daley:96,Parades:01} has revealed a highly sinuous pore network with a predominant and uniform nanometer size pore distribution ($\sim$90\% of the total pore volume of $\sim$0.4-0.5 cc/g determined from N$_2$ adsorption isotherms \cite{Daley:96,LiuW:14}). The remainder of the pore volume consists of largely random mesopores and some ultra-micropores. Various other measurements including thermodynamics\cite{Kaneko:92a, Kaneko:92b}, small-angle neutron and X-ray scattering \cite{Suzuki:92,Ramsay:98} of ACF-10 have also confirmed the interconnected pore structure of ACF-10, which is made primarily of elongated curvy slit-pores.  The pores themselves are just the voids or gaps between distorted (curled or curved, and not necessarily parallel) graphene sheets,  much like empty spaces within a dense pile of potato chips, but comparatively smaller \cite{Hayes:14}.   The key here is that the pores in ACF-10 are clearly not of the cylindrical-type as in the carbon or silica nanotubes, but rather of irregular slit-geometry, as evidenced in STM. This different pore structure offers a new platform for investigating fluids in confined geometries, in non-cylindrical shaped pores and evaluate the effect of pore geometry.  For the current ACF-10 sample, the micropore size distribution is $\sim$11.6$\pm$4 {\AA}, as determined from N$_2$ adsorption isotherms  \cite{Daley:96,LiuW:14} using the Dubnin-Astakhov equation \cite{Dubinin:89}. The nominal sample specific surface area is about 1000 m$^2$/g,  as confirmed at the time of manufacture by iodine number testing, which correlates well with BET calculations in this range \cite{Daley:96}. 

To prepare for the neutron experiments, we outgassed the as-provided ACF-10 sample for 36 hours at 473 K in a vacuum oven to remove all of the bulk-like water, and most of the surface water that was originally present in the as-received sample. We  then exposed approximately 2.3 grams of the \lq dried' batch to 99.9\% humidity in a desiccator for several hours. The hydration level reached after 24 hours was 21\%, as determined from the relative weight change of the sample. This rather important hydration level signals the possible presence of  hydrophilic groups (oxygenated) in an otherwise totally hydrophobic sample. Since this hydration amount is relative to the ' drying' conditions set above, the diffusive dynamics reported in this work are those of all water molecules present inside the porous carbon network.  The neutron being primarily sensitive to hydrogen atoms in QENS measurements, the measurements yield the characteristic relaxations of all confined water molecules.  The hydrated sample was thus carefully loaded within a 2 mm gap of two concentric Al cylinders to minimize multiple scattering. The subsequently indium-sealed containers were anchored to the copper finger of a close-cycle refrigerator (CCR) stick which allowed to control the sample temperature between 20 to 500 K, within $\pm$0.02 K.

\subsection{Elastic Incoherent Scattering}
We started our neutron measurements with a rapid collection of elastic scattering data  of the hydrated ACF sample, on cooling it from room temperature down to $\sim$100 K. The purpose of these scans were two-fold; (1) to investigate the absence of bulk-like water in the samples, and (2) to determine a temperature range where the dynamics can be suitably studied on the spectrometer. Fig. \ref{fig.1} shows the integrated elastic intensity $I(T)$ as a function of temperature for selected $Q$, normalized to the lowest temperature data collected at 100 K such that $I(T)/I(T_0)$ equals unity. The elastic intensity at each temperature was obtained by integrating the corresponding spectrum around its elastic peak, over a small energy range $\pm2\Gamma_r $, corresponding to the resolution width. Assuming an isotropic harmonic system, we can expect the elastic intensity to be proportional to $\exp{[-Q^2\langle r^2(T)\rangle/3]}$ \cite{Chen:99,Magazu:08}, where $\langle r^2(T)\rangle$ represents the mean square displacement associated with the hydrogen atoms in the water molecules. An increase in $\langle r^2(T)\rangle$ as a function increasing temperature  indicates an increased in mobility of the H-sites of water, and so of water itself. As a result, the elastic intensity  within  the 3.5 $\mu$eV energy resolution  would effectively increase with decreasing temperature until it reaches a maximum (plateau region at about 150 K). Crystallization, if present, would be observed as an abrupt change in $I(T)$, which is not seen here at any $Q$. This suggests that the adsorbed water does not solidify and remains mobile down to at least 150 K.

 Assuming that the diffusion process responsible for the observed drop in $I(T)$ has a unimodal and continuous diffusion (single Lorentzian), with temperature dependent width $\Gamma(T)\propto \exp(-E_A/RT)$, the normalized $I(T)$ at each $Q$ can  be fitted using the following expression \cite{Springer:77},

\begin{equation}
I(T)= A_0+(1-A_0)\arctan\left(\frac{\Gamma_r}{\Gamma(T)}\right)
\label{eq:EIS}
\end{equation}
\noindent where $\Gamma_r$ is the half-width of the instrument resolution, defined previously, and $A_0$ a fraction estimate of the observable molecules that do not participate in the diffusion process.  Fitted $E_A$ values are displayed in Table \ref{tab.2}. The solid lines in Fig. \ref{fig.1} represent the fits obtained, and corresponds to $E_A\simeq 14-15$ KJ/mol. 

\begin{figure}
\includegraphics[width=1.00\linewidth]{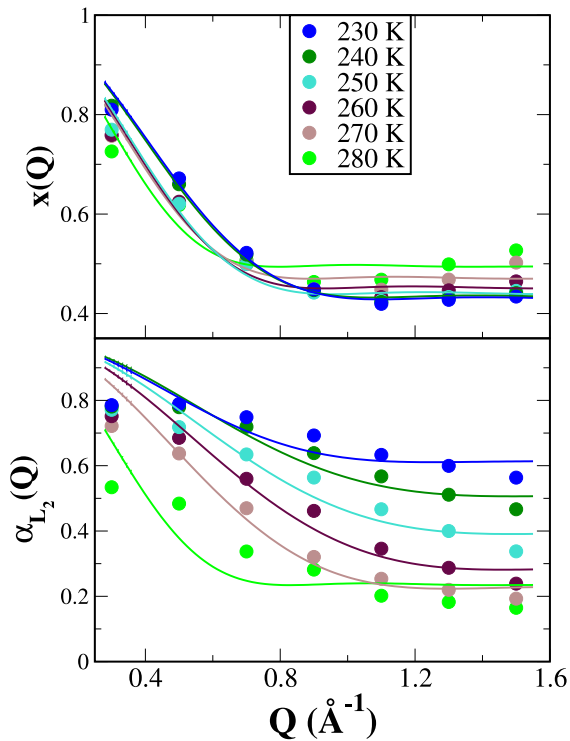}
\caption{(Color online) Plot of the fraction of elastic intensity ($x(Q)$ or elastic intensity structure factor ($EISF$)) (top panel) , and the relative weight of the narrowest Lorentzian $\alpha_{L_2}$ (bottom panel) as a function of momentum $Q$ and temperature $T$. Corresponding temperatures are shown in the legend. Colored solid lines are fits of the generic model for localized dynamics, discussed in the text. The $x(Q)$ parameter  is much more sensitive to $Q$ than it is to $T$, while $\alpha_{L_2}$ reveals a rather strong dependence on $Q$ and $T$.}
\label{fig.3}
\end{figure}

\begin{figure}
\includegraphics[width=1.00\linewidth]{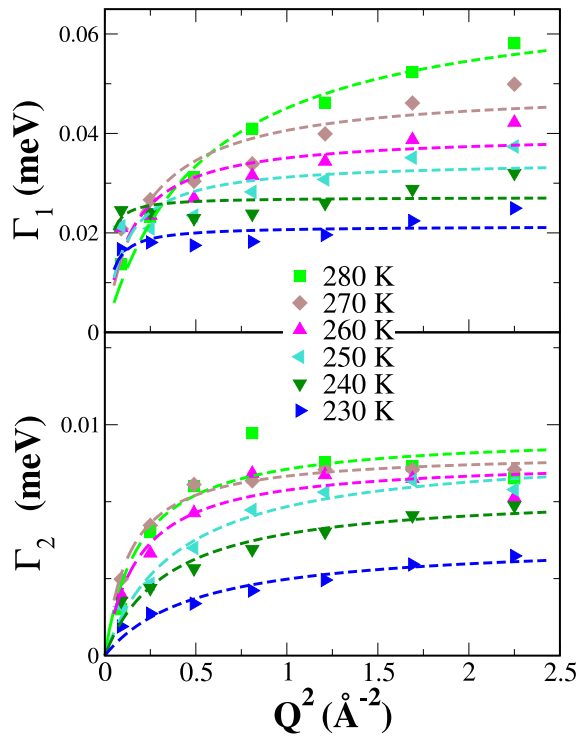}
\caption{(Color online) Temperature dependence of the Lorentzian linewidths ($\Gamma_1$ and $\Gamma_2$) as a function of $Q^2$. The solid symbols are experimental values, and the dashed lines  fits of the jump diffusion model. }
\label{fig.4}
\end{figure}

\begin{figure}
\includegraphics[width=1.00\linewidth]{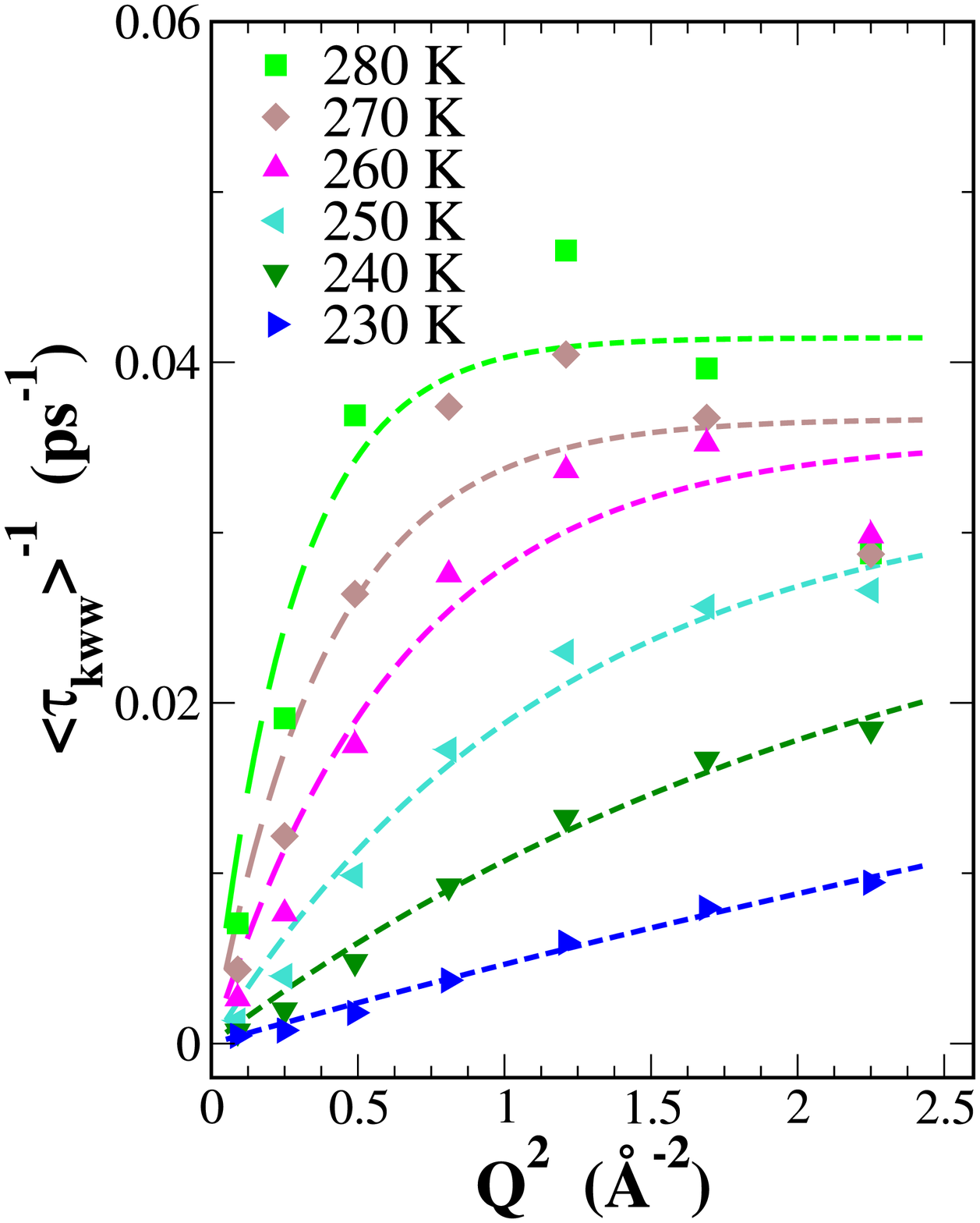}
\caption{(Color online) Inverse average relaxation time observed with a KWW model plotted as a function of $Q^2$ for 230 $\le T\le$ 280 K. The solid symbols are experimental values, and the dashed lines fits of the jump diffusion model.}
\label{fig.5}
\end{figure}

\subsection{Quasi-Elastic Scattering (QENS)}
Having determined a suitable temperature range for investigating the diffusion of water in ACF-10 on the spectrometer, we collected high statistical quality QENS data at six temperatures: 280, 270, 260, 250, 240 and 230 K for the hydrated ACF-10 sample.  The range of relevant momentum transfer for these QENS measurements  is 0.5 $\le Q\le$1.7  {\AA}$^{-1}$, $\Delta Q=0.2$ \AA$^{-1}$.  Fig. \ref{fig.2} depicts representative spectra at two temperatures for $Q=0.5$ \AA$^{-1}$, plotted against a spectrum obtained at 50 K, which is taken as the instrument resolution function. Below, we discuss our analysis  of the QENS data, from which we determine valuable parameters such as the diffusion coefficient, and characteristic relaxation time for the water's hydrogen sites, as a function of temperature.  

\begin{figure}
\includegraphics[width=1.0\linewidth]{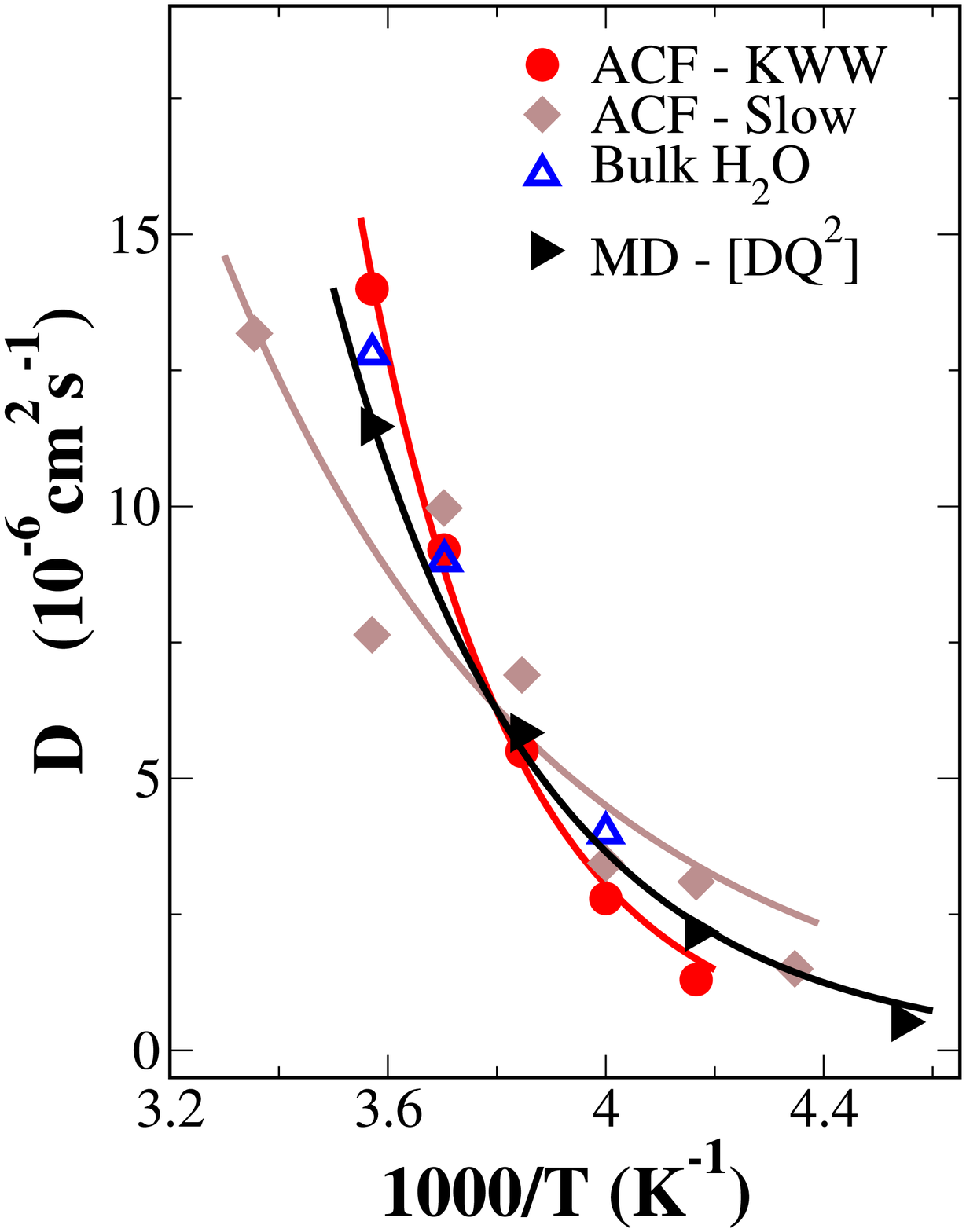}
\caption{(Color online) Average diffusion coefficient  as a function of $1000/T$. Other reference data are shown for comparison. Dashed lines are fits to an Arrhenius behavior $\sim D_0 e^{-E_a/RT}$}
\label{fig.6}
\end{figure}

\begin{figure}
\includegraphics[width=1.0\linewidth]{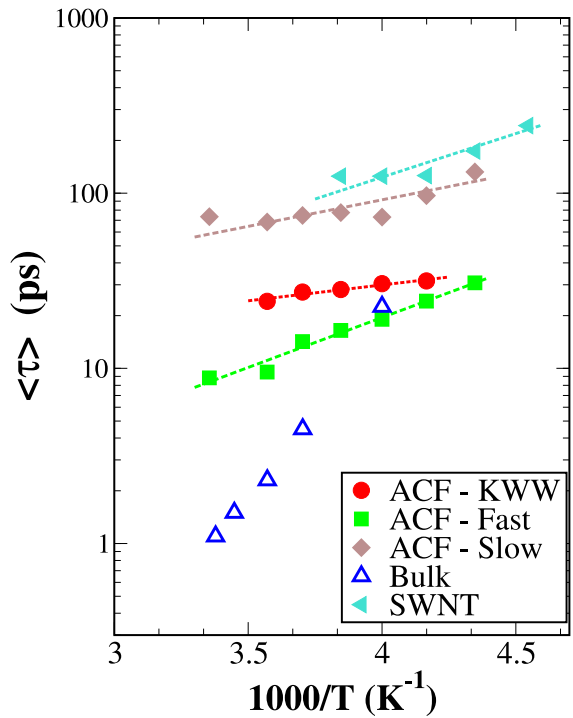}
\caption{(Color online) Average relaxation time of water confined in ACF-10 as a function of $1000/T$. Data for bulk water \cite{Teixeira:85,Takahara:05}, and for water confined in single-wall carbon nanotubes (SWNT)\cite{Mamontov:06} are shown for comparison. The molecular dynamics simulation (MD) results at 0.45 {\AA}$^{-1}$ fitted with the corresponding $KWW$ model are also shown.  Dashed lines are fits to an Arrhenius temperature dependence. For meaningful comparison with the SWNT, the ACF values are those obtained  using the same jump model in Ref. \onlinecite{Mamontov:06}.  Eq. \ref{eq.tauJump} applied to both data leads to somewhat lower values but similar temperature dependence.}
\label{fig.7}
\end{figure}

\begin{table}
\caption{Activation energy (rounded to the nearest integer) obtained from various sources. Values from fits of Eq. \ref{eq:EIS} to the $I(T)$ shown in Fig. \ref{fig.1}, and from the Arrhenius temperature dependent diffusion coefficients (Slow (2), KWW, MD, and bulk), as determined from Fig. \ref{fig.6} are shown.}
\label{tab.2}
\begin{ruledtabular}
\begin{tabular}{ c   c c c c c c}
Source      &$I(T)$  & $D_2$ & $D_{KWW}$ & MD & Bulk\\
\hline
$E_A$ (kJ/mol) & 15 & 14 & 29 & 22 & 24
\end{tabular}
\end{ruledtabular}
\end{table}

\subsection{Data Analysis } \label{sec:dan}
To analyze the experimental QENS data, we fitted it the observed spectra $S(Q,E)$ at each $Q$ and temperature point to a model function $S_m(Q,E)$ convoluted with the instrument resolution function $R(Q,E)$, plus a linear background term $B(Q,E)$;
\begin{equation}
S(Q,E)=S_m(Q,E)\otimes R(Q,E) +B(Q,E).
\end{equation}
\noindent Within the spectrometer energy window, our model $S_m(Q,E)$ function can be separated into two terms; an elastic part due to \lq static'  water molecules and a diffusive term $S_d(Q,E)$  coming from the dynamics of the molecules. If we denote the fraction of immobile water molecules by $x(Q)$ (commonly referred to as elastic incoherent structure factor or $EISF$), this model becomes, 
  \begin{eqnarray}
  S_{m}(Q,E)= x(Q)\times\delta(E)+(1-x(Q))\times S_{d}(Q,E)
 \label{eq.IQ}
 \end{eqnarray}
\noindent where $S_d(E)$ represents the diffusive component of the scattering. 

\section{Molecular Dynamics}\label{sec:MD}
Here we discuss the methodology employed in our molecular dynamics studies. Computer simulations coupled with QENS provide unparalled insights that are often impossible to obtain with experimental or theoretical approach alone. To further investigate the confinement and temperature effects on the water dynamics using molecular dynamics simulations, we study the time evolution of the intermediate scattering functions (ISF) for the water-hydrogen sites, $I_{in}(Q,t)=\langle \exp \left( iQ[{\bf r}_j(t_0+t)-{\bf r}_j(t_0)]\right)\rangle$, where ${\bf r}_j(t)$  is the position of the hydrogen-site $j$ at time $t$, $Q$ is the wavenumber (or momentum transfer), where $\langle \dots\rangle$  represents the averaging over all $j$-sites and time origins $t_0$ \cite{Chialvo:13,Chialvo:14}.  For that purpose we used a pair of flat grapheme plates, $L_x=28.40$ {\AA} by $L_y=30.74$ {\AA} , with an inter-plate separation $h=11.6$ \AA.  Each graphene plate comprised 364 carbon sites described atomistically as Lennard-Jones spheres characterized by   $\epsilon_{cc}/k=28$ K and $\sigma_{cc}/k=3.40$ {\AA}   \cite{Steele:73,Striolo:04}. These plates were immersed into an aqueous fluctuating cubic simulation box to perform isobaric-isothermal molecular dynamics at ambient pressure and temperatures in the range $220\le T \le 280$ K , according to a Nos\'e-Poincare symplectic integration algorithm \cite{Nose:01, Okumura:07} with a time-step of 2.0 fs, where the aqueous environment consists of 2048 SPC/E water molecules \cite{Berendsen:87}.  

According to the time-dependent configurational trajectories from simulation we calculated  $I_{in}(Q,t)$ at four $Q$-values, i.e., 1.80, 1.35, 0.90, and 0.45 {\AA}$^{-1}$ corresponding roughly to the intrinsic length-scales characterizing the system, within a time span of about 500 picoseconds. Additional details regarding the simulations are given elsewhere  \cite{Chialvo:13,Chialvo:14}.  To characterize the dynamics of confined water and its temperature response we regressed the simulated   $I_{in}(Q,t)$ in terms of either a two-exponential function,  $\alpha\exp(-t/\tau_1)+(1-\alpha)\exp(-t/\tau_2)$, or a stretched exponential function,  $\exp[-(t/\tau_s)^\beta]$, characterized by the relaxation time   $\tau_s$ and the stretching exponent  $\beta$.

\section{Results and Discussion}
To accurately represent all diffusive processes of water, the scattering model $S_d(Q,E)$ in Eq. \ref{eq.IQ} would ideally be a sum of several Lorentzian functions $\sum_i\frac{\Gamma_i}{E^2+\Gamma_i^2}$, each accounting for a specific diffusion $i$ in the sample. In practice however,  QENS data are most reliably characterized by no more than 2 Lorentzians (2L), still providing reasonable representation of all relevant diffusion processes. In general for water, the 2L model  works well for an instrument with broad dynamics range such as BASIS, and consists of a fast component, with characteristic width $\Gamma_1$, and a slow component with width $\Gamma_2$ . The fast component is generally associated with diffusion inside a \lq transient' cage (with say radius $a_1$) made of other water molecules, and the slow component with inter-cage diffusion \cite{Qvist:11}. In most instances however, whether the 2L is suitable or not,  a stretched exponential model (referred to as Kohlrausch-Williams-Watts (KWW) \cite{Chen:99,Li:12} from now on (i.e. $e^{-(t/\tau)^{\beta}}$ in the time domain or its Fourier-Transform in energy space) can always be used to represent the distribution of proton dynamics present in the sample.  We have used  both models to fit the data - not only for quantifying the dynamics, but also for determining the best model for our particular system.  We kept the stretching exponent $\beta$ fixed to 0.5  at all temperatures for convenience.  Our  goal was to systematically compare the average relaxation time and diffusion coefficient based on the two models. The other important aim was to compare the experimentally determined characteristic parameters with those obtained in previous experiments with single-wall nanotubes (SWNT), and from molecular dynamics simulations to which we return below. Interestingly, we find that the KWW model yields a better agreement between the simulated and measured diffusion coefficients $D$, despite the fact that both  models are able to reliably capture the experimentally observed values. In the 2L model however, $D$ could only be determined from the slower diffusive component, as generally found in confined systems \cite{Zanotti:99a}. We could not extract $D$ for the broad component because $\Gamma_1$ does not go to zero at low $Q$, as is expected in confined diffusion. Assuming a 2L model with Lorentzian components $L_1(Q,E)$ and $L_2(Q,E)$, $S_d(Q,E)$ can be written as,

\begin{equation}
S_d(Q,E)=(1-\alpha_{L_2}(Q))L_1(Q,E)+\alpha_{L_2}(Q)L_2(Q,E)
\label{eq.2L}
\end{equation}
\noindent where $\alpha_{L_2}(Q)$ is the relative weight of the narrow Lorentzian $L_2(Q,E)$. The $Q$-dependence of this parameter and that of the $x(Q)$ in Eq. \ref{eq.IQ}, which are displayed in Fig. \ref{fig.3}, can provide important details regarding the spatial extent of the confined diffusion in the ACF-10 sample.   We find an $x(Q)$ that varies little with temperature $T$  but noticeably with momentum transfer $Q$.  In contrast, the parameter $\alpha_{L_2}(Q)$ reveals a rather strong systematic dependence on both $Q$ and $T$. This allows to estimate the corresponding confining radius $a_1$ associated with the transient dynamical restriction by fitting  $\alpha_{L_2}$ to a generic  localized $EISF$ model $\left[ f + (1-f)\left(\frac{3j_1(Q a_i)}{Q a_i}\right)^2\right]$ \cite{Zanotti:99}. The corresponding fits are indicated by the solid lines in Fig. \ref{fig.3} for both $\alpha_{L_2}(Q)$ and $x(Q)$.  In the case of  $x(Q)$, the radius $a_0$ extracted yields the spatial confinement due to the nanopores in ACF-10, which is not  expected to vary significantly with temperature, as observed here. The fitted values are summarized in Table \ref{tab.1}.  Within a satisfactory fit uncertainty of 10 to 15\% ($\sim$ 0.3-0.7 \AA), we find a clear systematic reduction of the transient confining radius $a_1$ with decreasing temperature, while $f$ increases.

\begin{table}
\caption{Temperature dependence of the confining radius (in {\AA}) determined from model fit to the elastic parameter $x(Q)$, and the weight fraction of the second Lorentzian $\alpha_{L_2}$.}
\label{tab.1}
\begin{ruledtabular}
\begin{tabular}{c | c c c c c c c}
T (K) & 280 & 270 & 260 & 250 & 240 & 230 \\
\hline
$a_0 ({\AA})$ & 5.64 & 4.97 & 4.75& 4.70 & 4.80 & 4.10\\
$a_1$ ({\AA}) &  5.44 & 3.36  & 3.08  & 3.03 & 3.02 & 3.59
\end{tabular}
\end{ruledtabular}
\end{table}

Using Eq. \ref{eq.2L}, we have determined the $Q$-dependence of both $\Gamma_1$, and $\Gamma_2$, as summarized in Fig. \ref{fig.4}. The colored dashed lines  are the best model fits using the jump diffusion model,
\begin{equation}
\label{eq.tauJump}
\Gamma_i=\frac{D_i Q^2}{1+D\tau_{0i}Q^2}
\end{equation}
\noindent where $\tau_{0i}$ is the average residence time between jumps, and $D_i$ denotes the diffusion coefficient of process $i$. From these fits, we could reliably determine the characteristic relaxations for both processes, and the diffusion coefficient $D_2$ for the slower component. The convergence of $\Gamma_1$ to a plateau-like regime at low $Q$ complicates the determination of $D_1$.  

\begin{figure}
\includegraphics[width=1.00\linewidth]{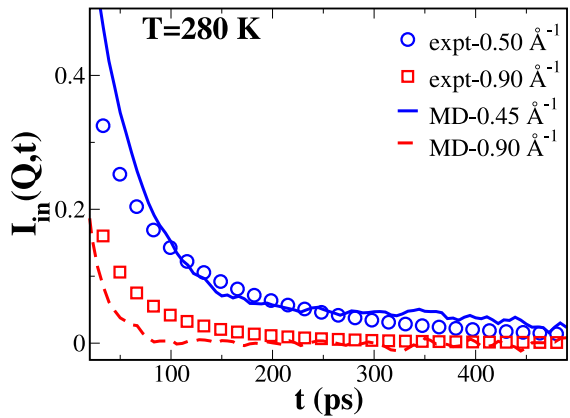}
\caption{(Color online) Simulated and calculated intrinsic intermediate scattering function $I_{in}(Q,t)$ of the confined water molecules at 280 K, and selected momentum transfer $Q$.  The measured decay component  is $I_{in}(Q,t)=e^{-(t/\tau_{KWW})^{\beta}}$ where $\tau_{KWW}$ and $\beta$ are the best fit parameters to the experimental data using  Eq. \ref{eq.IQ} with the KWW model.}
\label{fig.8}
\end{figure}

\begin{figure}
\includegraphics[width=1.00\linewidth]{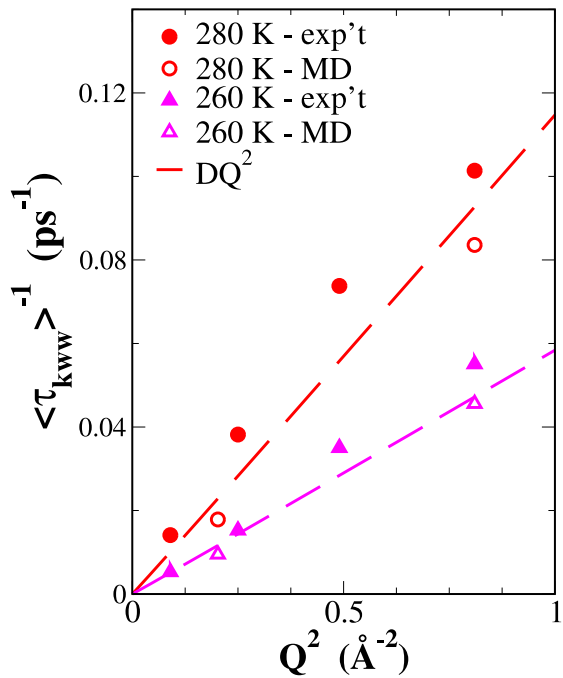}
\caption{(Color online) Comparison of the simulated and measured $\langle \tau\rangle^{-1}$ at low $Q$. The average relaxation times shown were derived from the KWW model for consistency. The dashed lines are quadratic $Q^2$ fits to the simulated data.}
\label{fig.9}
\end{figure}

We have also evaluated the diffusion characteristics with the alternative KWW model discussed above. The extracted  average relaxation times are plotted as a function of $Q^2$ are shown in Fig. \ref{fig.5}. The dashed lines are fitted line using Eq. \ref{eq.tauJump}.  The fits however reveal different behavior. Specifically, the KWW fits indicate a diffusion process that goes from a jump-type diffusion at ambient temperature to a continuous one below 250 K. For this reason, we only compare the average relaxation time $\langle \tau_{KWW}\rangle$ with those obtained using the 2L model for $T$ between 250 K and 280 K. The extracted diffusion coefficients ($D_{KWW}$ and $D_2$) and average relaxation time $\langle\tau\rangle$ on the other hand are compared at all temperatures. These comparison are summarized in Figs. \ref{fig.6} and \ref{fig.7}. Data from bulk water \cite{Teixeira:85,Takahara:05} and SWNT \cite{Mamontov:06} are displayed for comparison, as well as the results from our molecular dynamics at low $Q$.  While the observed KWW diffusion parameter $D_{KWW}$ is a bit smaller than that reported for bulk water, the $D_2$ or slow diffusion coefficient for the narrow Lorentzian in the 2L model, is significantly smaller. This is not unreasonable since the slow process may be overshadowed in the KWW model by other faster processes. However, we find a much better agreement between the calculated and simulated $D$ with the KWW model, as evidenced in Fig. \ref{fig.6}.  

Since a direct comparison between the simulated $Q$-dependent $\langle \tau_Q\rangle$  and the experimentally observed $Q$-independent $\langle \tau\rangle$ ( high $Q$-limit of the measured $\langle \tau_Q\rangle$ modeled with the jump diffusion model in Eq. \ref{eq.tauJump}) is not meaningful, the $Q$-dependent MD data are not shown in Fig. \ref{fig.7}. Instead, we chose to directly compare the simulated and measured $I(Q,t)$ and the corresponding $Q$-dependent  $\langle \tau_Q\rangle$ at each temperature.  To do this, we note that the simulations provide the intrinsic water dynamics while the neutron data contains both the dynamics and structure component. In order to adequately compare the intrinsic intermediate scattering function $I_{in}(Q,t)$, we show in Fig. \ref{fig.8} the simulated $I(Q,t)$ along with only the experimentally observed stretched decay component at 280 K (i.e. $e^{-(t/\tau_{KWW})^{\beta}}$ where $\tau_{KWW}$ is the fitted experimental value -- omitting the $EISF$ term). Reasonable agreement can be found between the two, particularly at the lowest $Q$ value.  We find some discrepancies in the faster decay at low times which would require further investigations. The long times show generally a reasonable agreement at these low $Q$  values,  which unfortunately degrades as $Q$ increases.   To highlight the level of agreement  at low $Q$ ( $\le$ 1.0 \AA$^{-1}$), we compare the simulated and experimentally observed $\langle \tau_{KWW}\rangle^{-1}$ in Fig. \ref{fig.9}.  The dashed lines are the corresponding $DQ^2$ fits from which the $D$ values from molecular dynamics (MD) (shown Fig. \ref{fig.6})  were determined.

\begin{figure}
\includegraphics[width=1.00\linewidth]{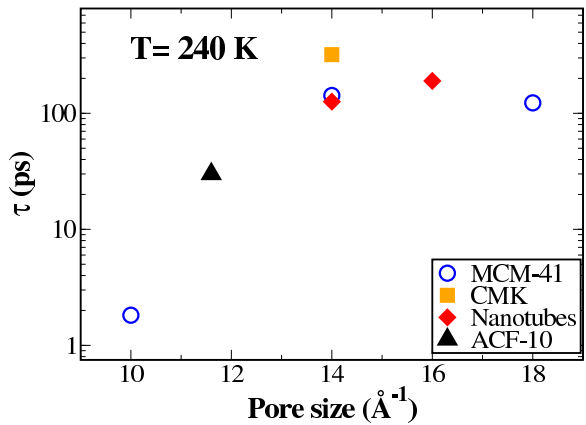}
\caption{(Color online) Average relaxation time $\langle \tau \rangle$ of water confined in various microporous materials as a function of confining pore size at selected temperature of 240 K. The open symbols denote values observed for the hydrophilic silica-based (MCM-41S) materials in Ref. \cite{Liu:06}, and the filled symbols are for hydrophobic carbon-based materials, respectively highly ordered cabon molecular sieves CMK \cite{Chu:10}, cylindrical carbon nanotubes \cite{Mamontov:06}, and our current activated carbon ACF-10 fibers. }
\label{fig.10}
\end{figure}

\section{Conclusions} \label{sec:conc}

In summary, we have investigated the diffusion of water confined in a network of narrow wavy slits in activated carbon fiber by means of quasi-elastic neutron, and molecular dynamics. By  comparing the experimentally observed characteristic diffusion coefficients and relaxation times using the same analysis approach, we conclude that the diffusion of water in ACF-10 is slower than in bulk water but  of comparable magnitude to that in cylindrical carbon SWNT (only marginally smaller)  and other confining porous media of comparable pore size. Our main conclusion is that above 1 nm, it is not the geometry of the confining pores, but rather their dimensions that primarily affect the dynamics of confined water above 1 nm.  Stated differently, the diffusive dynamics of water in pores of the same size above 10 {\AA}, irrespective of whether they are cylindrical or not (slit or not) appears to be of the same order of magnitude for the temperature range investigated, as is illustrated for example in Fig. \ref{fig.10} for temperature $T=$240 K.   Data at other temperatures revealed similar behavior. To validate (or disprove) this perhaps fortuitous observation and evaluate its implication, future experimental work should systematically compare the effect of pore size on water dynamics in the complete series of ACF made from the same percursor (ACF-15, ACF-20 and ACF-25), to that of comparable hydrophilic silica samples (MCM-41 or other), all modeled using the same approach, since $\langle \tau\rangle$ values are known to be model-dependent. We expect however that MD simulations of water in hydrophobic and hydrophilic pores of varying pore size and geometry would yield predictive behavior.  A previous experimental report \cite{Chathoh:11} in which $\langle\tau\rangle$ values obtained from various samples using different models  (KWW or 2L) were compared, found no particular dependency of $\langle\tau\rangle$  on pore size.

The experimentally observed relaxation dynamics of confined water are supported by molecular dynamics  simulations of an ideal scenario in which water is restricted to diffuse within  the interstitial gap between two parallel graphene layers while in equilibrium with its aqueous bulk.  The simulations yield a reasonable agreement at low momentum transfer $Q$ ($Q\le1.0$ \AA${^{-1}}$).  The origin of the discrepancies at larger $Q$ remains to be clarified but could arise from either the relatively simple model used, the importance of localized dynamics or rotational motions, the presence of  oxygen-containing groups on the graphene surface or edge (carbonyl or hydroxyl for example).  Localized dynamics are strongly influenced by surface effects such as surface roughness \cite{Chialvo:14}, disorder and corrugation, which becomes dominant at high $Q$. The simulations yield however a diffusion coefficient $D$ that is in excellent agreement with the experimental value.  The comparison between QENS and simulation suggests the need for the additional interrogation of the effect of confined morphology, such as corrugation and pore connectivity, in order to aid the interpretation of the observed discrepancies in the relaxation times.

\acknowledgments
We thank R. Moody (SNS) for her help with the sample preparation, and R. Goyette (SNS) for his technical assistance at the beamline. This research was supported as part of the Fluid Interface Reactions, Structures and Transport (FIRST) center, an Energy Frontier Research Center funded by the U.S. Department of Energy, Office of Basic Energy Sciences.  Research at ORNL's SNS is sponsored by the Scientific User Facilities Division, Office of Basic Energy Sciences, U.S. Department of Energy.  The TEM measurements and analysis were performed at the ORNL's Center for Nanophase Materials Sciences, which is a DOE Office of Science User Facility.

{\footnotesize
$^*$\lq Kynol$^{TM}$ is a registered trademark of Gun Ei Chemical Industry Co., Ltd. for novoloid fibers and textiles.

\end{document}